\documentclass{aa}  

\usepackage{graphicx}
\usepackage{txfonts}
\usepackage{amsmath}
\usepackage{amsfonts}
\usepackage{natbib}
\usepackage{longtable}
\usepackage{lscape}
\usepackage[usenames, dvipsnames]{xcolor}
\usepackage[normalem]{ulem}
\usepackage{bm}
\usepackage{esvect}
\usepackage{multirow}
\usepackage{varwidth}
\usepackage{array}
\usepackage{subfig}

\DeclareMathOperator{\Tr}{Tr}

\begin{document} 

    \title{The star formation history of Upper Scorpius and Ophiuchus}
  \subtitle{A 7D picture: positions, kinematics, and dynamical traceback ages\footnote{A Table with the data collected in this study, including new radial velocities and the membership to the sub-populations identified in this study, is only available in electronic form at the CDS via anonymous ftp to \url{cdsarc.u-strasbg.fr} (130.79.128.5) or via \url{http://cdsweb.u-strasbg.fr/cgi-bin/qcat?J/A+A/}} }

   \author{N.~Miret-Roig\inst{1}
          \and
          P.~A.~B.~Galli\inst{2}
          \and
          J.~Olivares\inst{3}
          \and
          H.~Bouy\inst{4}
          \and
          J.~Alves\inst{1}
          \and
          D. Barrado\inst{5}
          }

   \institute{University of Vienna, Department of Astrophysics, Türkenschanzstraße 17, 1180 Wien, Austria\\
   e-mail: \url{nuria.miret.roig@univie.ac.at}
         \and
         Núcleo de Astrofísica Teórica, Universidade Cidade de São Paulo, R. Galvão Bueno 868, Liberdade, 01506-000, São Paulo, SP, Brazil.
         \and
         Departamento de Inteligencia Artificial, Universidad Nacional de Educación a Distancia (UNED), c/Juan del Rosal 16, E-28040, Madrid, Spain.
         \and
         Laboratoire d'astrophysique de Bordeaux, Univ. Bordeaux, CNRS, B18N, allée Geoffroy Saint-Hilaire, 33615 Pessac, France.
         \and
         Centro de Astrobiología (CAB), CSIC-INTA, ESAC Campus, Camino Bajo del Castillo s/n, 28692 Villanueva de la Cañada, Madrid, Spain.
             }

   \date{Received ; accepted}

  \abstract
   {Understanding how star formation begins and propagates through molecular clouds is a fundamental but still open question. One major difficulty in addressing this question is the lack of precise 3D kinematics and age information of young stellar populations. Thanks to \textit{Gaia}'s astrometry, large spectroscopic surveys, and improved age-dating methods, this picture is changing.}
   {We aim to study spatial and kinematic substructures of the region encompassed by Upper Scorpius and Ophiuchus star forming regions. We want to determine dynamical traceback ages and study the star formation history of the complex.}
   {We combined our spectroscopic observations with spectra in public archives and large radial velocity surveys to obtain a precise radial velocity sample to complement the \textit{Gaia} astrometry. We used a Gaussian Mixture Model to identify different kinematic structures in the 6D space of positions and velocities. We applied an orbital traceback analysis to estimate a dynamical traceback age for each group and determine the place where it was born.}
   {We identified seven different groups in this region. Four groups ($\nu$~Sco, $\beta$~Sco, $\sigma$~Sco and $\delta$~Sco) are part of Upper Scorpius, two groups ($\rho$~Oph and $\alpha$~Sco) are in Ophiuchus, and another group ($\pi$~Sco) is a nearby young population. We found an age gradient from the $\rho$~Oph group (the youngest) to the $\delta$~Sco group ($\lesssim 5$~Myr), showing that star formation was a sequential process for the past 5~Myr. Our traceback analysis shows that Upper Scorpius and $\rho$~Oph groups share a common origin. The closer group of $\pi$~Sco is probably older, and the traceback analysis suggests that this group and the $\alpha$~Sco group have a different origin, likely related to other associations in the Sco-Cen complex.}
   {Our study shows that this region has a complex star formation history that goes beyond the current formation scenario, likely a result of stellar feedback from massive stars, supernovae explosions, and dynamic interactions between stellar groups and the molecular gas. In particular, we speculate that photo-ionisation from the massive $\delta$~Sco star could have triggered star formation first in the $\beta$~Sco group and then in the $\nu$~Sco group. The perturbations of stellar orbits due to stellar feedback and dynamical interactions could also be responsible for the 1--3~Myr difference that we found between dynamical traceback ages and isochronal ages. }
  

   \maketitle
%

\section{Introduction}\label{sec:intro}

The Upper Scorpius OB stellar association is the youngest group of the Scorpius-Centaurus association \citep{deZeeuw99}.
The region occupied by Upper Scorpius (USC) and Ophiuchus (Oph), which are part of the association, constitutes one of the closest (140~pc) star forming complexes and therefore, constitutes an excellent laboratory to investigate how star formation begins and propagates through molecular clouds. \citet{Preibisch+1999} proposed that star formation in Upper Scorpius was triggered by a supernova explosion in Upper Centaurus-Lupus around 5~Myr ago and that 1.5~Myr ago the most massive star in Upper Scorpius exploded as a supernova, triggering star formation in Ophiuchus. This scenario builds on previous work pointing out that Ophiuchus is affected by the feedback of massive stars in Upper Scorpius \citep[e.g.][]{Elmegreen+1977, Loren1989, Wilking+2008}, and is consistent with a recent analysis of the motion of the Sco-Cen association \citep{Zucker+2022}. However, there is still an ongoing debate on the age of Upper Scorpius that could upend the currently accepted formation scenario for the region.

Despite its proximity to Earth, it is notable that there is currently no consensus on the age of Upper-Scorpius. Several studies find a young age for Upper-Sco of around 5~Myr \citep{deGeus+1989, Preibisch+2002, Slesnick+2008, David+2019, Asensio-Torres+2019} while, others find older values of around 10~Myr \citep{Pecaut+2012, Rizzuto+2016, Feiden2016}. All these studies used evolutionary models which depend on the included physics. This age uncertainty has important implications on many studies of star formation, such as the mass function \citep{Miret-Roig+2022} and the study of disc and planet formation \citep{Barenfeld+2016, Esplin+2018, Richert+2018}. Some studies suggested that age discrepancies could be caused by a spread in age in the region or by limitations and uncertainties on evolutionary models \citep{Herczeg+2015, Feiden2016, Fang+2017, Asensio-Torres+2019, Sullivan+2021} but there is not yet a consensus on the age of Upper Scorpius. 

Recently, many authors have studied the substructure of the region using the 5D astrometry of \textit{Gaia} \citep[see e.g.][]{Damiani+2019, Kerr+2021, Squicciarini+2021, Luhman2022, Ratzenbock+2022, Briceno-Morales+2022}. They all agree that Upper Scorpius is not a single population of stars but there is, instead, a rich substructure of different young populations. However, different authors have identified different numbers of groups and group members, indicating that the different populations are perhaps related but likely highly overlapping, in particular in 2D projection. 

The \textit{Gaia} mission \citep{GaiaCol+16} has measured the 3D spatial distribution and 2D tangential velocities of many stars with precision of the order of parsecs and few hundreds of meters per second, respectively. To perform the most detailed study of this region, we need to complement the \textit{Gaia} astrometry with radial velocities of a precision similar to the tangential velocities. The \textit{Gaia} Data Release 3 (DR3, \citealt{GaiaColVallenari+2022}) has provided the community with the largest sample of homogeneous radial velocities to date, however, still the most precise radial velocities come from ground-based surveys like APOGEE \citep{Majewski+2017}, which measured hundreds of thousands of radial velocities with a typical uncertainty of a few hundreds of meters per second. 

In this work, we aim to revisit the spatial structure, kinematics, and star formation history of the Upper Scorpius and Ophiuchus complex. The extra value of our work compared to previous studies is two-fold: 1) We used a new dataset, combining the 5D astrometry of \textit{Gaia} plus the ground-based radial velocities of APOGEE complemented with our own and archival observations and \textit{Gaia} DR3. 2) These measurements were then fed to a new methodology to determine dynamical traceback ages that we presented in \cite{Miret-Roig+2020b}, representing the best opportunity to study this region in 7D (3D positions, 3D velocities and age) with unprecedented precision. This article is structured as follows. In Section~\ref{sec:data} we describe our initial sample and our search for the most precise 6D phase space data, in Section~\ref{sec:substructure} we describe the algorithm we used to identify the substructure in the sample, in Section~\ref{sec:ages} we describe the methodology we used to obtain a dynamical traceback age for each group, in Section~\ref{sec:SFH} we propose a new star formation history for this complex, and in Section~\ref{sec:conclusions} we present our conclusions.

\section{Data}\label{sec:data}

Our initial sample contains 2\,812 members of Upper Scorpius and Ophiuchus, selected with \textit{Gaia} DR2 or \textit{Hipparcos} astrometry from \citet{Miret-Roig+2022}\footnote{We included the $\rho$~Oph and $\chi$~Oph, B-type stars, which are well known members of the young association but not identified as members by the study because of their binary nature and extinction. They both have membership probabilities greater than 0.5 but below the threshold established in that study to avoid a large fraction of contaminants.}. In this section we describe the astrometry and radial velocities that we compiled to obtain the phase space positions of individual stars. Table~\ref{tab:num_sources} indicates the number of objects at each step of the sample selection process. An electronic Table with the data collected in this study, including new radial velocities and the results of the clustering analysis (Sect.~\ref{sec:substructure}), will be available at the CDS.

\begin{table}
   \begin{center}
    \caption{Number of sources at each step of the data selection process (see Sect.~\ref{sec:data}).}
    \label{tab:num_sources}
    \begin{tabular}{l|rr}
    \hline
    \hline
    {Members from \citet{Miret-Roig+2022}}           & 2\,812 &  \\
    {\qquad \textit{Gaia} DR3 astrometry}      & 2\,805 &  \\
    \hline                      
    {Radial velocity}                                &    Total & Precise$^*$   \\
    {\qquad this work}                               &    157 &  98 \\
    {\qquad \textit{Gaia} DR3}                       &   1\,103 &  99 \\
    {\qquad APOGEE DR17}                             &    991 & 934 \\ 
    \hline                      
    {6D data (\textit{Gaia} astrometry + RV)}        &        & \\
    {\qquad Suspected SB }                           &     89 & \\ 
    {\qquad Single}                                  &    981 & \\
    {\qquad Single + RV filtering}                   &    871 & \\
    \hline     
    {Bona fide sample }                               &          & \\
    {\qquad 5D}                                       &  2\,190 &  \\
    {\qquad 6D phase space}                           &     670 &  \\
    \hline
    \hline
    \end{tabular}
    \end{center}{}
    \noindent\tablefoottext{*}{With a precision of <1~km~s$^{-1}$.} 
\end{table}{}

\subsection{Proper motions and parallaxes}

We matched our initial sample with the \textit{Gaia} DR3 catalogue using the 2D position (RA and Dec with a maximum separation of 1\arcsec). We found a counterpart for all sources in our initial sample except for $\alpha$~Sco (Antares) and $\delta$~Sco. Three other OB stars ($\beta$~Sco, $\sigma$~Sco and $\tau$~Sco) are in \textit{Gaia} DR3 but do not have a complete astrometric solution. For these five stars, we used the parallax and proper motions from \textit{Hipparcos} \citep{vanLeeuwen+2007}. Two other stars (RX~J1600.5--2027 and 2MASS~J16015149--2445249) had an astrometric solution in \textit{Gaia} DR2 but not in \textit{Gaia} DR3 and were excluded from the rest of analysis.
The median uncertainties of the sample are around 0.06~mas~yr$^{-1}$ in proper motions, 0.05~mas in parallax and 0.14~km~s$^{-1}$ in tangential velocity.

\subsection{Radial velocities}

Radial velocities are an essential parameter to study kinematics in 3D. In this section, we describe the catalogue of radial velocities that we obtained from our spectroscopic observations plus spectra in public archives and how we combined it with large radial velocity surveys. 

\subsubsection{Radial velocities measured in this work}

We observed 63 targets of our sample with the CHIRON spectrograph at the SMARTS 1.5~m telescope (P.I. Bouy, NOIRLab Programs 2020A-0094 and 2021A-0011) and 7 targets with the HERMES spectrograph at the 1.2~m Mercator telescope (P.I. Barrado, Program 5-Mercator1/21A).
To complement our observations, we searched for spectra of our targets in the European Southern Observatory (ESO) and Haute-Provence Observatory (OHP) public archives. Table~\ref{tab:num-spectra} provides an overview of the number of spectra analysed with different instruments. We downloaded and processed all the available spectra to obtain radial velocities with the same methodology as in \citet{Miret-Roig+2020b}. We obtained a radial velocity measure for 157 sources (available in the electronic Table) with a median uncertainty of 0.7~km~s$^{-1}$. From these, 20 stars do not have a radial velocity in the large surveys considered in this study (\textit{Gaia} DR3 and APOGEE DR17). This value increases to 46 if we consider only radial velocities with a precision better than 1~km~s$^{-1}$ in the large surveys.

\begin{table}
\centering
\caption{Archival spectra analysed in this study. 
\label{tab:num-spectra}}
\begin{tabular}{lrr@{$-$}rr}
\hline\hline
  \multicolumn{1}{c}{Spectrograph} &
  \multicolumn{1}{c}{$R$} & 
  \multicolumn{2}{c}{$\Delta\lambda$} & 
  \multicolumn{1}{c}{\# Spectra} \\
  
  \multicolumn{1}{c}{} &
  \multicolumn{1}{c}{} & 
  \multicolumn{2}{c}{(nm)} & 
  \multicolumn{1}{c}{} \\
\hline\hline
GIRAFFE   &  $20\,000$ & $370$\,  & $900$   & 460   \\
FEROS     &  $48\,000$ & $350$\,  & $920$   & 285   \\
UVES      & $110\,000$ & $300$\,  & $1100$  & 201   \\
HARPS     & $115\,000$ & $378$\,  & $691$   & 108   \\
CHIRON    &  $80\,000$ & $410$\,  & $870$   &  80   \\
HERMES    &  $85\,000$ & $380$\,  & $900$   &   7   \\
ELODIE    &  $45\,000$ & $385$\,  & $680$   &   2   \\
\hline\hline
\end{tabular}
\tablefoot{The total number of spectra analysed is 1\,138 and some sources have multiple spectra. The (maximum) resolving power and spectral range of each spectrograph are indicated.}
\end{table}

\subsubsection{Combining our radial velocities with large surveys}

\begin{figure}
    \centering
    \includegraphics[width = \columnwidth]{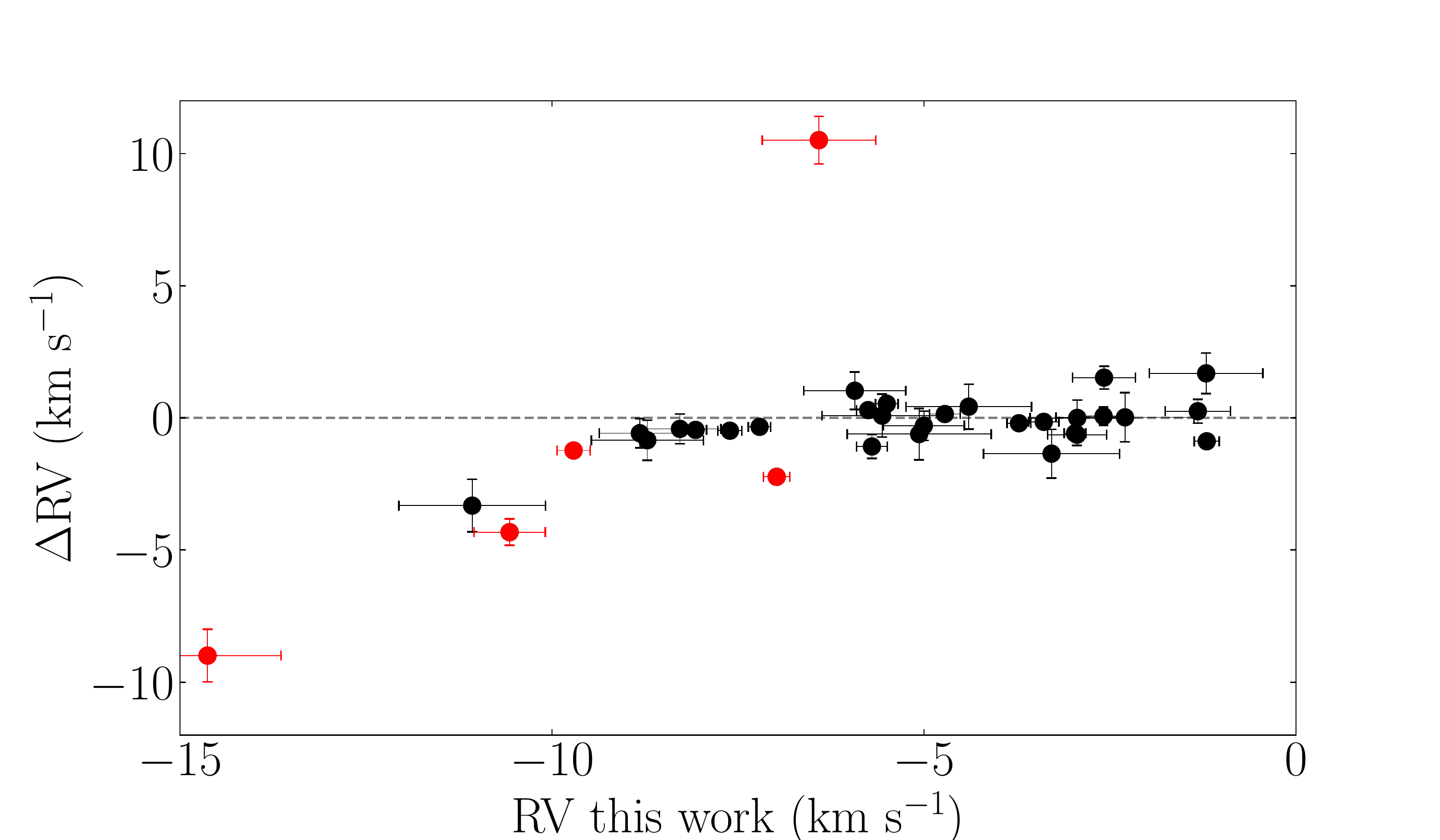}
    \includegraphics[width = \columnwidth]{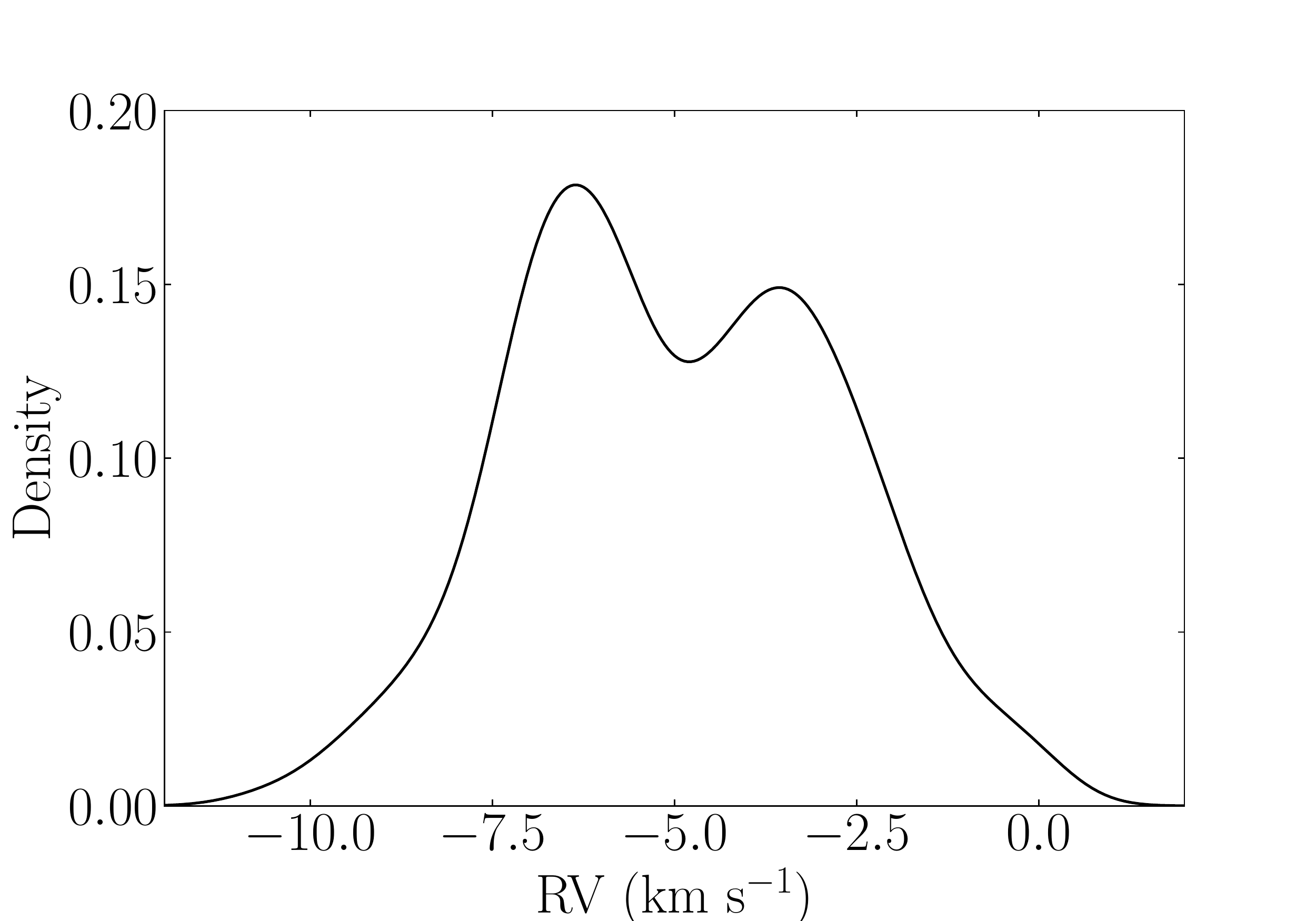}
    \caption{Top: Radial velocity residuals between this work and APOGEE ($\Delta$RV = RV this work -- RV APOGEE). The radial velocity measures which are not compatible within $5\sigma$ uncertainties are indicated in red. Bottom: Radial velocity distribution of the bona fide sample. We used a Gaussian kernel density with a bandwidth of 0.25~km~s$^{-1}$, similar to the radial velocity precision of the sample.}
    \label{fig:rv_residuals}
\end{figure}

We complemented our own radial velocity measurements with the two largest homogeneous radial velocity catalogues to date, covering this region, APOGEE DR17 \citep{Abdurrouf+2021} and \textit{Gaia} DR3 \citep{Katz+2022}. To combine the radial velocities in the three catalogues (the one we presented in this section, APOGEE and \textit{Gaia}) into our final sample, we only considered measurements with a precision below 1~km~s$^{-1}$. This is the minimum precision necessary to obtain dynamical traceback ages of young associations with reasonable accuracy \citep{Miret-Roig+18} and similar to the typical dispersion of young associations. Additionally, we set a minimum uncertainty of 0.1~km~s$^{-1}$ whenever there was a smaller value. This threshold is similar to the precision in tangential velocities and facilitates the study of the kinematic substructure (see Sect.~\ref{sec:substructure} and \citealt{Miret-Roig+2020b}). We obtained the final radial velocity measurement as the weighted average\footnote{We used the inverse of the squared uncertainty as weights ($w_i=(\sigma_{RV, i})^{-2}$) and the uncertainty in the weighted average as $\sigma_{RV, WA} = [\Sigma_i w_i]^{-1}$. } of the three catalogues: this work, APOGEE, and \textit{Gaia}. 

The radial velocities of spectroscopic binaries are not useful for a traceback analysis unless the velocity of the centre of masses is known, which is not the case in general. Therefore, we identified all known binaries and set their radial velocities as missing values. We identified 26 stars which have more than one component in APOGEE, 18 stars with an entry in one of the \textit{non-single stars} tables of \textit{Gaia} DR3, and 38 stars where classified as suspected binaries in our study for having double-peaked or very broad cross-correlation functions (some of these were also classified as binaries in other studies \citealt{David+2019, Stauffer+2018, Rosero+2011, Ratzka+2005, Perez+2004}). These stars are flagged as binaries in our electronic catalogue.

In Figure~\ref{fig:rv_residuals} (top) we show the difference between the radial velocities measured in this study and those in APOGEE, for the 35 stars that have a precise measure in both catalogues with an error smaller than 1~km~s$^{-1}$. The known binaries have been excluded from this comparison. We have noted that six sources have radial velocity measures that are not compatible within 5$\sigma$ uncertainties, among the two catalogues. This is an indicator that they could be spectroscopic binaries, and thus their radial velocity has also been set as missing value.

The final distribution of radial velocities is multivariate since it includes several kinematic populations (see Figure~\ref{fig:rv_residuals}, bottom). Despite the relatively wide spread of the population, several sources have a radial velocity measurement which is clearly not consistent with this region. Since our sample has been selected with extremely precise \textit{Gaia} astrometry and photometry, we expect a very low contamination rate ($\lesssim1$\%, \citealt{Miret-Roig+2022}). Therefore, we have defined a radial velocity filtering criterion to exclude outliers in the radial velocity distribution. We kept only the radial velocity measures in the 90\% central distribution, and discarded the radial velocities that fall outside the percentiles $p_5$, $p_{95}$. These 108 sources are candidates to be spectroscopic binaries (with not enough observations to be detected) or problematic measures.

\subsection{Bona fide sample}\label{subsec:bona-fide} 

To identify the substructure of this region and obtain precise dynamical traceback ages, we defined a subsample of stars with the most precise data. We only considered sources classified as members by \citet{Miret-Roig+2022} using  \textit{Gaia} astrometry. We applied the filtering criteria described in \citet{Fabricius+2021} to identify potential non-single objects and $5\sigma$ outliers in parallax or proper motion. These criteria are based on the renormalised unit weight error (RUWE) parameter (\texttt{ruwe} $<1.4$) and two image-level indicators (\texttt{ipd\_gof\_harmonic\_amplitude}$ <0.1$ and \texttt{ipd\_frac\_multi\_peak} $\leq 2$). After this filtering, the sample contains 2\,190 sources (78\% of the initial sample) with 5D \textit{Gaia} astrometry and 670 sources with 6D phase space data.

\section{Groups in the 6D phase space}\label{sec:substructure}

\begin{figure*}
    \centering
    \includegraphics[width = \textwidth]{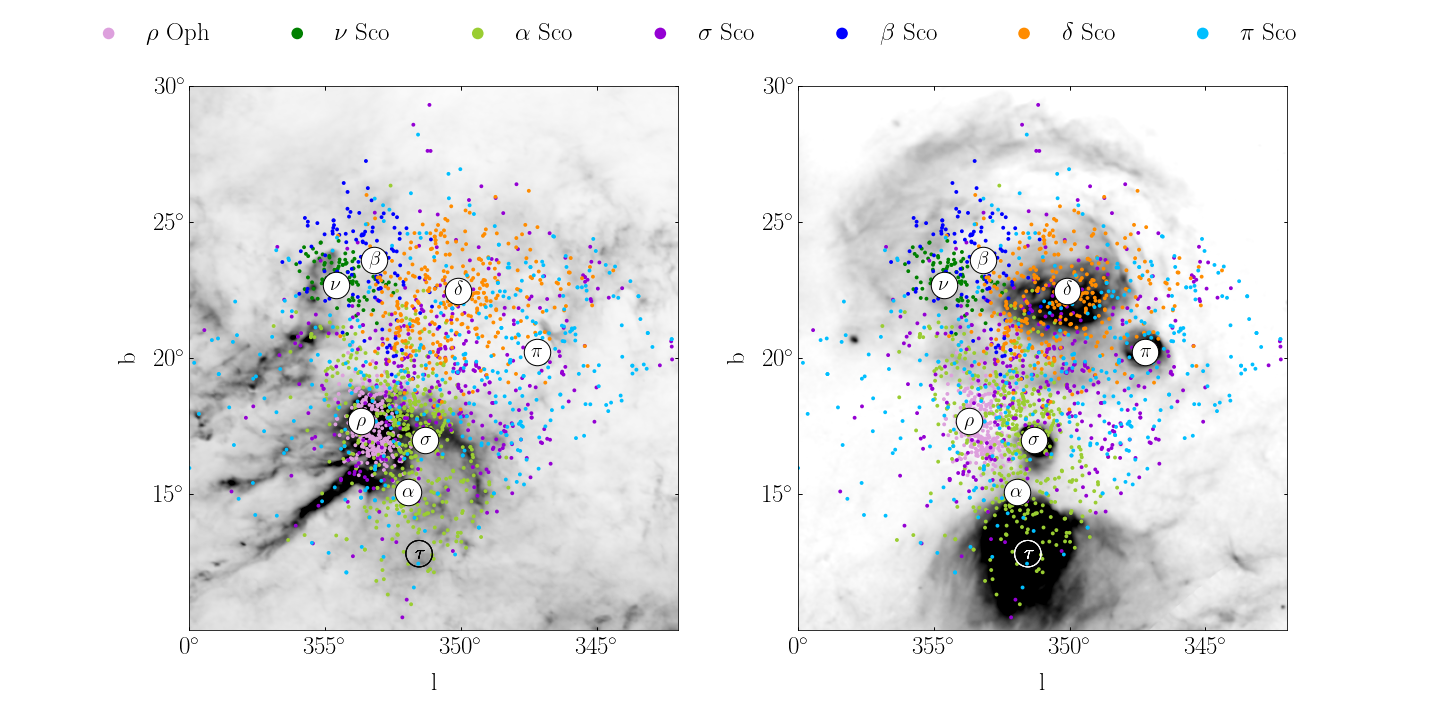}
    \caption{Distribution of the groups in the plane of the sky in galactic coordinates. The position of the B-type stars giving name to the groups are indicated by the white circles. The location of the $\tau$~Sco star, discussed in the text, is shown. The background image shows the emission in radio, at 857 GHz \citep[left, ][]{Planck+2020} and in H$\alpha$ \citep[right, ][]{Finkbeiner+2003}. }
    \label{fig:sky_position} 
\end{figure*}

\begin{table*}
\centering
\caption{Groups identified in the region of Upper Scorpius and Ophiuchus. 
\label{tab:groups}}
\begin{tabular}{l|ccc|ccc|ccc|ccc|ccc}
\hline\hline
  \multicolumn{1}{c|}{Group} &
  \multicolumn{3}{c|}{Num. members} & 
  \multicolumn{1}{c}{X} & 
  \multicolumn{1}{c}{Y} & 
  \multicolumn{1}{c|}{Z} & 
  \multicolumn{1}{c}{U} & 
  \multicolumn{1}{c}{V} & 
  \multicolumn{1}{c|}{W} &
  \multicolumn{1}{c}{$\sigma_X$} & 
  \multicolumn{1}{c}{$\sigma_Y$} & 
  \multicolumn{1}{c|}{$\sigma_Z$} & 
  \multicolumn{1}{c}{$\sigma_U$} & 
  \multicolumn{1}{c}{$\sigma_V$} & 
  \multicolumn{1}{c}{$\sigma_W$} \\
  
  \multicolumn{1}{c|}{} &
  \multicolumn{1}{c}{Total} & 
  \multicolumn{1}{c}{BF 5D} & 
  \multicolumn{1}{c|}{BF 6D} & 
  \multicolumn{3}{c|}{(pc)} & 
  \multicolumn{3}{c|}{(km~s$^{-1}$)} & 
  \multicolumn{3}{c|}{(pc)} & 
  \multicolumn{3}{c}{(km~s$^{-1}$)} \\ 

\hline\hline
$\rho$ Oph   &      415 &      339 &       94 &     131.7 &     -15.7 &      41.4 &      -5.9 &     -15.1 &      -9.2 &    3.7 &    1.3 &    2.6 &    1.6 &    0.9 &    1.2 \\
$\nu$ Sco    &      140 &      110 &       57 &     127.5 &     -12.4 &      54.2 &      -5.7 &     -15.4 &      -8.8 &    2.6 &    1.9 &    1.7 &    1.0 &    0.5 &    1.1 \\
$\alpha$ Sco &      601 &      517 &      135 &     141.8 &     -20.3 &      44.7 &      -3.6 &     -16.5 &      -5.9 &   14.5 &    4.9 &    6.5 &    1.0 &    0.7 &    0.6 \\
$\sigma$ Sco &      539 &      357 &      104 &     141.0 &     -22.5 &      50.7 &      -5.6 &     -16.5 &      -7.5 &   13.4 &    7.7 &    9.1 &    2.2 &    1.5 &    1.3 \\
$\beta$ Sco  &      181 &      143 &       61 &     139.4 &     -16.3 &      61.3 &      -3.1 &     -16.1 &      -6.8 &    4.4 &    2.8 &    3.7 &    1.1 &    0.5 &    0.7 \\
$\delta$ Sco &      424 &      333 &      131 &     129.5 &     -21.9 &      53.3 &      -6.2 &     -16.1 &      -7.7 &    3.6 &    4.7 &    4.0 &    0.8 &    0.6 &    0.8 \\
$\pi$ Sco    &      510 &      391 &       88 &     117.5 &     -20.3 &      41.9 &      -4.7 &     -18.2 &      -4.6 &   16.9 &    8.6 &    7.6 &    1.8 &    0.9 &    1.0 \\
\hline\hline
\end{tabular}
\tablefoot{Number of members and properties of the groups identified in this study. Columns indicate: (1) the name of the group used in this study, (2) the total numbers of stars per group, (3) the number of bona fide stars with 5D astrometry per group, (4) the number of bona fide stars with 6D (astrometry and radial velocities) per group, (5--7) the median positions, (8--10) the median velocities, (11--13) the standard deviation of the positions, (14--16) the standard deviation in the velocities. }
\end{table*}

\textit{Gaia} has demonstrated that the spatial and kinematic structure of associations and star-forming regions is highly complex. In particular, several studies have recently analysed the region of Upper Scorpius finding different groups in this region \citep[e.g.][]{Damiani+2019, Kerr+2021, Squicciarini+2021, Ratzenbock+2022, Briceno-Morales+2022}. The phase space (\textit{XYZUVW}) is the best space to unravel the spatial and kinematic complexity since it avoids problems like projection effects. The main drawback of this space is that it requires accurate radial velocities, which are not available in the same quantity and quality as the 5D astrometry. Working in the phase space with limited radial velocities can then introduce some biases due to the incompleteness of the sample. For instance, the \textit{Gaia} radial velocities are magnitude limited and the APOGEE radial velocities are spatially limited with the footprint of the survey. The novelty of our study with respect to previous ones is that it searches for substructures in a 6D space in which we can include the sources with missing radial velocity by marginalising over the missing information. This is possible in a representation space that includes the radial velocity and not a transformation of it. This allows us to work in a complete 6D space and include a large fraction of the members in the region at the same time. Our methodology also accounts for the observational uncertainties. 

We used a Gaussian Mixture Model (GMM) in the 3D Galactic Cartesian positions, 2D Galactic tangential velocities and radial velocities space to classify all the members of Upper Scorpius and Ophiuchus into different substructures. We infer the parameters of this GMM using the algorithm developed by \citet[ see their Sect. 2.1.1]{Olivares+2018}. This algorithm has the advantage that it also uses the sources with missing radial velocity to find the best model of the observations, minimising any possible observational bias. First, we fitted the GMM to the 2\,190 sources in the 5D bona fide sample (30\% of which have a radial velocity measure in our final catalogue, excluding sources with large uncertainties and binaries, see Table~\ref{tab:num_sources} and Sect.~\ref{sec:data}). Once the best final model is established, we classified the complete sample of sources. 

We explored all models with a number of components between 1 and 17. We decided to work with the seven-component solution, which is the one with minimum Bayesian Information Criteria (BIC). We also analysed the rest of models and we found that, in more complex models, the new components are wider and tend to catch sources in the outskirts of the distribution of parameters in the representation space, which conform less reliable populations. We computed the solution for each number of components with five different seeds and in all cases we find groups with similar weights, means and covariance matrix. We repeated the same procedure in a different representation space, \textit{XYZUVW}, with similar results. The disadvantage of this space is that we cannot include the sources with missing information in the radial velocity, therefore, we prefer the first analysis. In Figure~\ref{fig:sky_position}, we show the distribution of the groups projected on the plane of the sky, where we also indicate the position of the giant star that gives name to the group. Several of these B-type stars were initially classified in the group of $\sigma$~Sco, which is very extended in the position space. However, the astrometry of these very bright stars is less precise and we reclassified them into their closest group in the plane of the sky. In Figure~\ref{fig:rs}, we show the distribution of groups in the representation space and in Table~\ref{tab:groups}, we summarise the mean properties of the groups we found.

The dispersion in the positions in the radial direction ($\sigma_X$) is slightly larger than in the other two directions. This is not surprising since the radial direction is the closest to the line of sight. The median uncertainty in the radial ($X$) positions is 1~pc, more than twice the median uncertainty in the azimuthal ($Y$) and vertical ($Z$) positions. The three components of the Cartesian velocities have similar precisions of few hundreds of meters per second due to the filtering in radial velocities that we applied (see Sect.~\ref{sec:data}). The dispersion in positions of the groups $\alpha$~Sco, $\sigma$~Sco and $\pi$~Sco is larger than the rest, indicating that these groups might be older populations which started to disperse into the Galactic field or that they were born sparser \citep{Ward+2020, Wright+2022}. The groups $\sigma$~Sco and $\pi$~Sco occupy the entire area analysed in this work, suggesting that we might be seeing only part of them. This could have implications in the determination of the dynamical traceback ages as we discuss in the following sections. The group of $\sigma$~Sco has the largest dispersion in velocities, which could indicate that this group is slightly more contaminated.

\subsection*{Comparison with previous studies}

We have cross-matched our list of members with the three recent studies that also identified several spatial and kinematic groups in this region, namely \citet{Kerr+2021}, \citet{Squicciarini+2021}, and \citet{Ratzenbock+2022}. In Table~\ref{tab:literature} we relate the names of the groups in our study to the names given in previous works. We have identified the groups in previous studies that share the majority of members with our groups, but there are always small differences in the list of members of each group. The groups of $\rho$~Oph, $\nu$~Sco, $\beta$~Sco, and $\delta$~Sco generally agree well in all studies. This gives a strong validation to these groups, since each study used different methodologies and representation spaces. On the contrary, the more dispersed groups of $\alpha$~Sco, $\sigma$~Sco, and $\pi$~Sco seem more mixed and different authors have separated them differently. Therefore, the results of these three groups should be considered with a bit more caution.

\section{Dynamical traceback ages}\label{sec:ages}

\begin{table}
\setlength{\tabcolsep}{3pt}
\centering
\caption{Dynamical traceback age summary. 
\label{tab:ages}}
\begin{tabular}{l|ccccc}
\hline\hline
  \multicolumn{1}{c|}{Group}  & 
  \multicolumn{1}{c}{Determinant} & 
  \multicolumn{1}{c}{Trace} & 
  \multicolumn{1}{c}{X} & 
  \multicolumn{1}{c}{Y} & 
  \multicolumn{1}{c}{Z} \\
 
  \multicolumn{1}{c|}{} &
  \multicolumn{1}{c}{(Myr)} & 
  \multicolumn{1}{c}{(Myr)} &
  \multicolumn{1}{c}{(Myr)} & 
  \multicolumn{1}{c}{(Myr)} & 
  \multicolumn{1}{c}{(Myr)} \\

\hline\hline
$\rho$ Oph   &       0.0 $\pm$  0.2 &   0.0 $\pm$ 0.3 &   0.1 $\pm$ 0.4 &   0.0 $\pm$ 0.3 &   0.0 $\pm$ 0.5 \\
$\nu$ Sco    &       0.2 $\pm$  0.5 &   0.3 $\pm$ 0.5 &   0.4 $\pm$ 0.6 &   0.5 $\pm$ 0.8 &   0.3 $\pm$ 1.0 \\
$\alpha$ Sco &       0.3 $\pm$  0.7 &   1.0 $\pm$ 1.2 &   1.3 $\pm$ 1.9 &   3.0 $\pm$ 1.5 &   0.2 $\pm$ 0.5 \\
$\sigma$ Sco &       2.4 $\pm$  0.6 &   2.1 $\pm$ 0.7 &   1.8 $\pm$ 1.1 &   2.7 $\pm$ 0.6 &   2.1 $\pm$ 1.0 \\
$\beta$ Sco  &       2.4 $\pm$  1.7 &   2.5 $\pm$ 1.6 &   2.7 $\pm$ 1.5 &   1.5 $\pm$ 1.1 &   3.9 $\pm$ 2.6 \\
$\delta$ Sco &       4.6 $\pm$  0.6 &   4.6 $\pm$ 1.1 &   1.6 $\pm$ 1.6 &   5.5 $\pm$ 0.7 &   3.8 $\pm$ 0.7 \\
$\pi$ Sco    &       5.1 $\pm$  2.1 &   6.3 $\pm$ 1.4 &   8.4 $\pm$ 2.9 &   2.8 $\pm$ 2.4 &   3.1 $\pm$ 1.6 \\

\hline\hline
\end{tabular}
\tablefoot{This table displays the dynamical traceback ages obtained with the different size estimators presented in Sect.~\ref{sec:ages}. }
\end{table}

\begin{figure}
    \centering
    \includegraphics[width = \columnwidth]{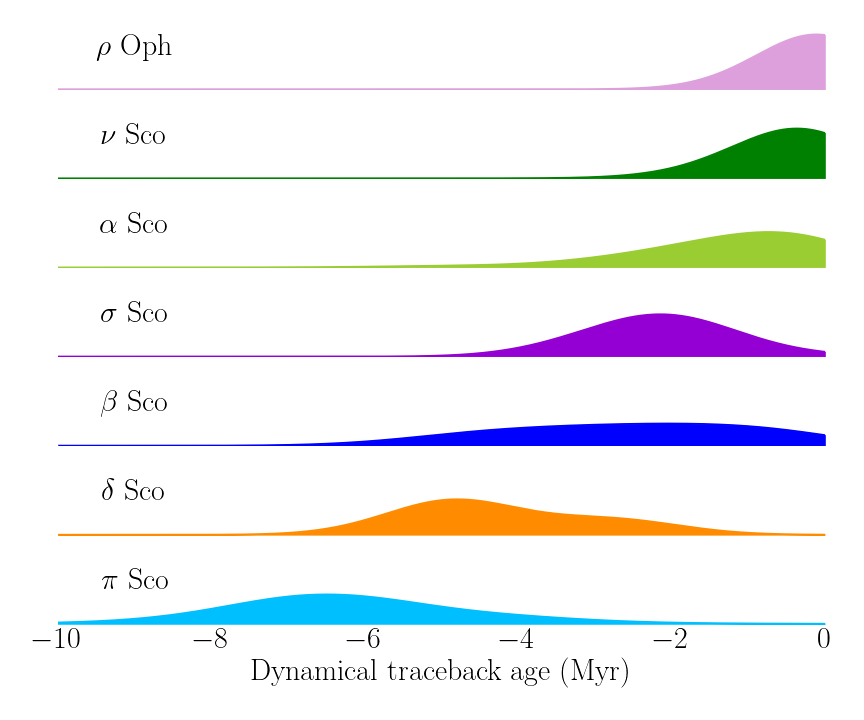}
    \caption{Dynamical traceback age distribution obtained for each of the groups using the Trace as the size estimator.}
    \label{fig:summary_age}
\end{figure}

We used the methodology described in \cite{Miret-Roig+18, Miret-Roig+2020b} to compute the dynamical traceback age for each of the groups identified in this study. In short, this methodology uses the present 3D positions and 3D velocities of individual stars and computes the stellar orbits back in time with a Galactic potential. We used two different 3D Milky Way potentials namely, \textit{MWPotential14} \citep{Bovy15} and \textit{McMillan17} \citep{McMillan17} and obtained the same results. We did the orbital integration with the \texttt{galpy} software\footnote{http://github.com/jobovy/galpy} \citep{Bovy15}. We refer to \citet{Miret-Roig+2020b} for more details on the impact of Galactic potentials on the dynamical traceback age of young stellar associations.

We define the dynamical traceback age as the time when a group of stars was most concentrated in the past, i.e. when the size of the group was minimal. We measured the size of a group of stars with a robust estimate of the covariance matrix ($\mathbf{\Sigma}$, the Minimum Covariance Determinant from Sklearn \citealt{scikit-learn}), which reduces the impact of outliers in the sample, and the three functions defined in \citet{Miret-Roig+2020b}. 
\begin{itemize}
    \item The size in the radial, azimuthal, and vertical directions ($S_{x}$, $S_{y}$, $S_{z}$) are the squared root of the diagonal terms of the covariance matrix in each direction.
    \item The Trace Covariance Matrix Size ($S_{TCM}$) is defined as: 
    \begin{equation}
        S_{TCM}  =  \left[ \frac{ \Tr (\mathbf{\Sigma})}{3} \right]^{1/2}.
        \label{eq:TCM}
    \end{equation}
    \item The Determinant Covariance Matrix Size ($S_{DCM}$) is defined as:
    \begin{equation}
        S_{DCM} =  \left[ \det (\mathbf{\Sigma}) \right]^{1/6}. 
        \label{eq:DCM}
    \end{equation}
\end{itemize}

We show the size of each group as a function of time using these three functions in Figure~\ref{fig:dyn_age_size_time} (left panels). We computed the dynamical traceback age with a 1\,000 bootstrap repetition with replacement, taking each time the same number of sources as the number of members in the group. This strategy accounts for the uncertainty in the dynamical traceback age from contamination in group members, which dominates the observational uncertainties and the uncertainties on different modern models of the Galactic potential \citep{Miret-Roig+2020b}. For each bootstrap repetition, we measured a dynamical traceback age which is shown in the age distributions in Figure~\ref{fig:dyn_age_size_time} (right panels). In some cases, the age distributions are wide or bi-modal indicating that there might be a degree of contamination in the sample. 

We measured a dynamical traceback age for each group using the five size estimators ($S_{x}$, $S_{y}$, $S_{z}$, $S_{TCM}$, $S_{DCM}$). In each case, we report the median and standard deviation of the dynamical traceback age distribution in Table~\ref{tab:ages}. The ages obtained with the determinant and the trace contain the full 3D information and are compatible within the uncertainties. Therefore, we use the dynamical traceback age obtained from the trace for the rest of the discussion, similar to what we did in \citet{Miret-Roig+2020b}. In Figure~\ref{fig:summary_age}, we show the dynamical traceback age distribution for each of the groups. We observe an age gradient among different groups, that is statistically significant, suggesting that star formation has been a sequential process, with several star formation bursts, for the past 5~Myr. Similar age gradients have been observed in other nearby star forming regions, like in Orion \citep{Mathieu+2008, Grossschedl+2021}, Vela \citep{Cantat-Gaudin+2019, Armstrong+2022}, Perseus \citep{Pavlidou+2021} and Cygnus \citep{Quintana+2022}. 

Recently, \citet{Squicciarini+2021} also measured kinematic traceback ages\footnote{We refer to their ages as \textit{kinematic} traceback ages because they did not use a Galactic potential to trace back the orbits of stars. We note that they use the term `kinematic ages' to refer to their results.} of different groups in Upper Scorpius. The main differences between their and our method are: i) they report the ages obtained with a 4D analysis (RA, Dec, tangential velocity in RA, Dec; see their Table 4, although they did some tests in 6D), ii) they used a different strategy to measure the size of a group, and iii) they did not use a Galactic potential to compute the orbits of stars back in time. For the groups where we have similar membership lists ($\rho$~Oph, $\nu$~Sco, $\beta$~Sco and $\delta$~Sco) we find results that are compatible within the uncertainties. The main difference is that in our study, the $\delta$~Sco group is 2~Myr older than the $\beta$~Sco group. 

\subsection*{Validation of the age gradient with isochrones and possible origin of a zero point shift} 

\begin{figure}
    \centering
    \includegraphics[width = \columnwidth]{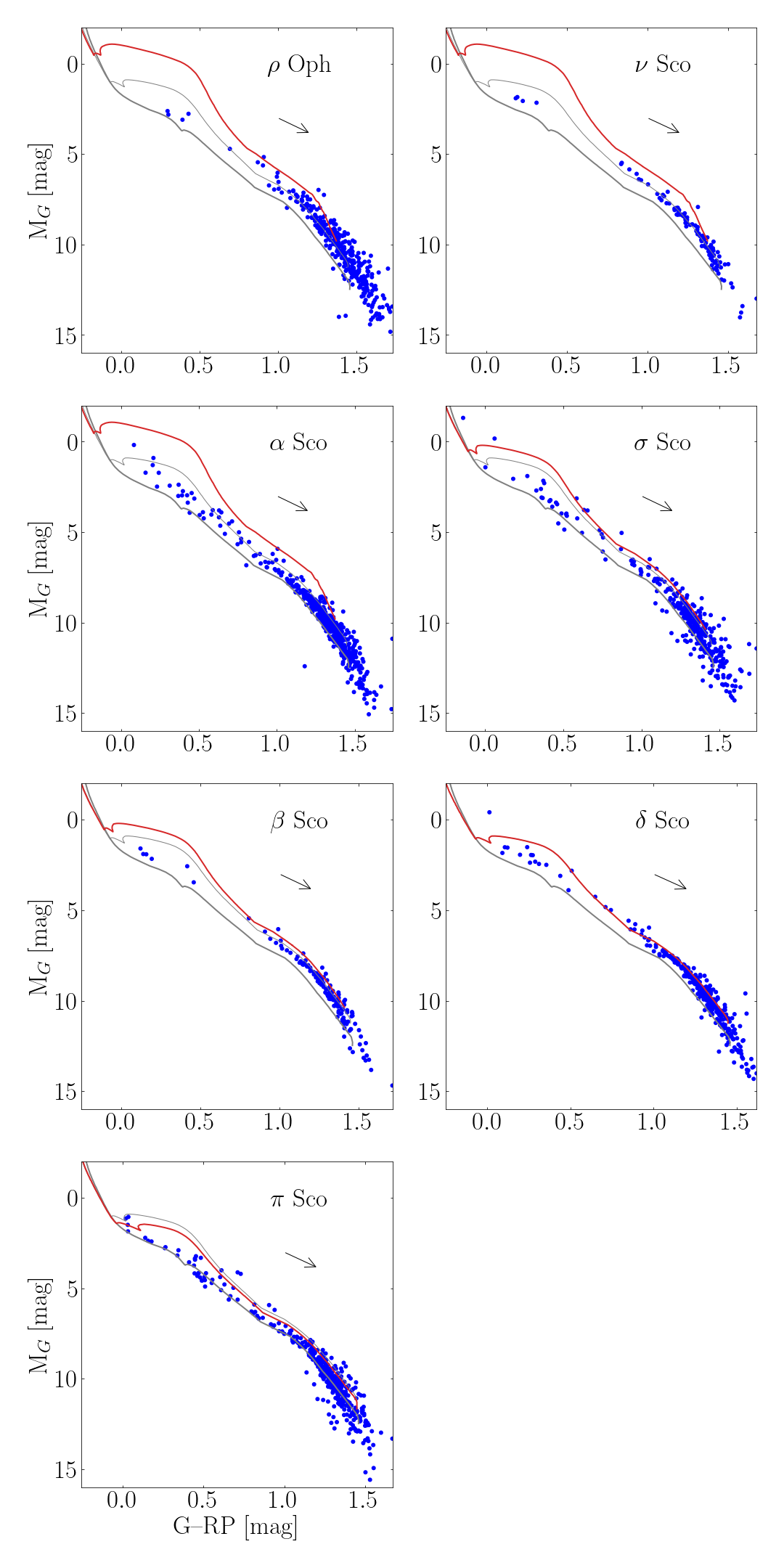}
    \caption{Colour magnitude diagram of the members of the groups identified in this study. The PARSEC isochrone corresponding to the dynamical traceback ages determined in this study are indicated in red. The PARSEC isochrones at 5 and 20 Myr and the 1~mag extinction vector are also indicated.}
    \label{fig:cmd}
\end{figure}

A common technique to determine stellar ages is the use of evolutionary models, which has proven to be very successful for intermediate and old clusters \citep[see eg.][]{Bossini+2019, Cantat-Gaudin+2020}. However, very young stars ($<20$~ Myr) are often variable (due to activity) and affected by interstellar extinction (which changes spatially in this region, see Fig.~\ref{fig:sky_position}). These and other effects make photometric ages for such young objects notoriously difficult to estimate (e.g. \citealt{Jeffries+2014, Jeffries+2021}, \citealt{Binks+2022}). Additionally, evolutionary models are specially uncertain at very young ages, making isochrone fitting less ideal for those cases \citep{Barrado+2016}. Determining isochronal ages from evolutionary models is beyond the scope of this work, but we still use colour-magnitude diagrams (CMDs) to validate the relative age scale determined in this section (Fig.~\ref{fig:summary_age}).

In Figure~\ref{fig:cmd}, we show the CMD of the seven groups identified in this work on top of the PARSEC isochrone  \citep{Marigo+17}, corresponding to the dynamical traceback ages determined in this study. A visual inspection shows that $\rho$~Oph and $\nu$~Sco are the youngest groups. The groups of $\alpha$~Sco, $\sigma$~Sco, $\beta$~Sco and $\delta$~Sco have intermediate ages and the scatter in the CMDs makes it difficult to determine relative differences between these four groups. The most likely reasons for this scatter are extinction, which is spatially variable (see Fig.~\ref{fig:sky_position}, left panel), and stellar variability. However, a slight contamination of the membership could also be responsible for part of the scatter. Finally, the group of $\pi$~Sco is the oldest in this region according to the CMD. The relative age scale obtained in this study from dynamical traceback ages and the one observed in the CMD generally agree within the uncertainties.

Looking at Figure~\ref{fig:cmd}, we see that the isochronal ages of these groups might be a few Myr older than the dynamical traceback ages determined in this study. This is especially true for $\pi$~Sco, which could have an age of around 20~Myr according to evolutionary models. However, this association likely extends beyond the area covered by our study, and our spatial selection could bias the dynamical traceback age. To explain the smaller age differences in the rest of groups we believe that gravitational interactions between different groups, stellar feedback, and the dissipation of the parent gas must have an important role. These effects are not accounted for in our study and can produce perturbations in the orbits of individual stars which would affect the estimated dynamical traceback ages. We estimate that these perturbations should be of the order of the age of the $\rho$~Oph group (1--3~Myr, \citealt{Greene+1995}) for which we could not determine a dynamical traceback age. Adding this 1--3~Myr offset to the dynamical traceback ages obtained for the rest of the groups reconciles the ages measured in this study with isochronal age measurements in this region. However, we add a last word of caution since isochronal ages for young stellar associations are also affected by other limitations as we have discussed at the beginning of this section.

Another technique used to measure kinematic ages consists of measuring the correlation between the velocity and position of all  members in a group \citep{Torres+06, Mamajek14, Miret-Roig+2020b}. We fitted a linear relation between the Cartesian heliocentric positions ($XYZ$) and velocities ($UVW$) independently to each of the three directions (Galactic centre, Galactic rotation and Galactic north pole). Although we found signs of expansion in all groups, we were unable to determine precise expansion ages for any group. We interpret this as most groups being too young to have undergone a significant expansion that can be measured in the present with this simple method.

\section{Star formation history}\label{sec:SFH}

In this section, we investigate the star formation history of the different groups identified in this study. First, we compute the orbit of the centre of each group using a 3D Galactic potential. Then, we discuss possible sources of feedback that could have triggered star formation in this region and perturbed the trajectories of stars. 

\subsection{Orbital traceback analysis}

\begin{figure*}
    \centering
    \sidecaption
    \includegraphics[width = 0.7\textwidth]{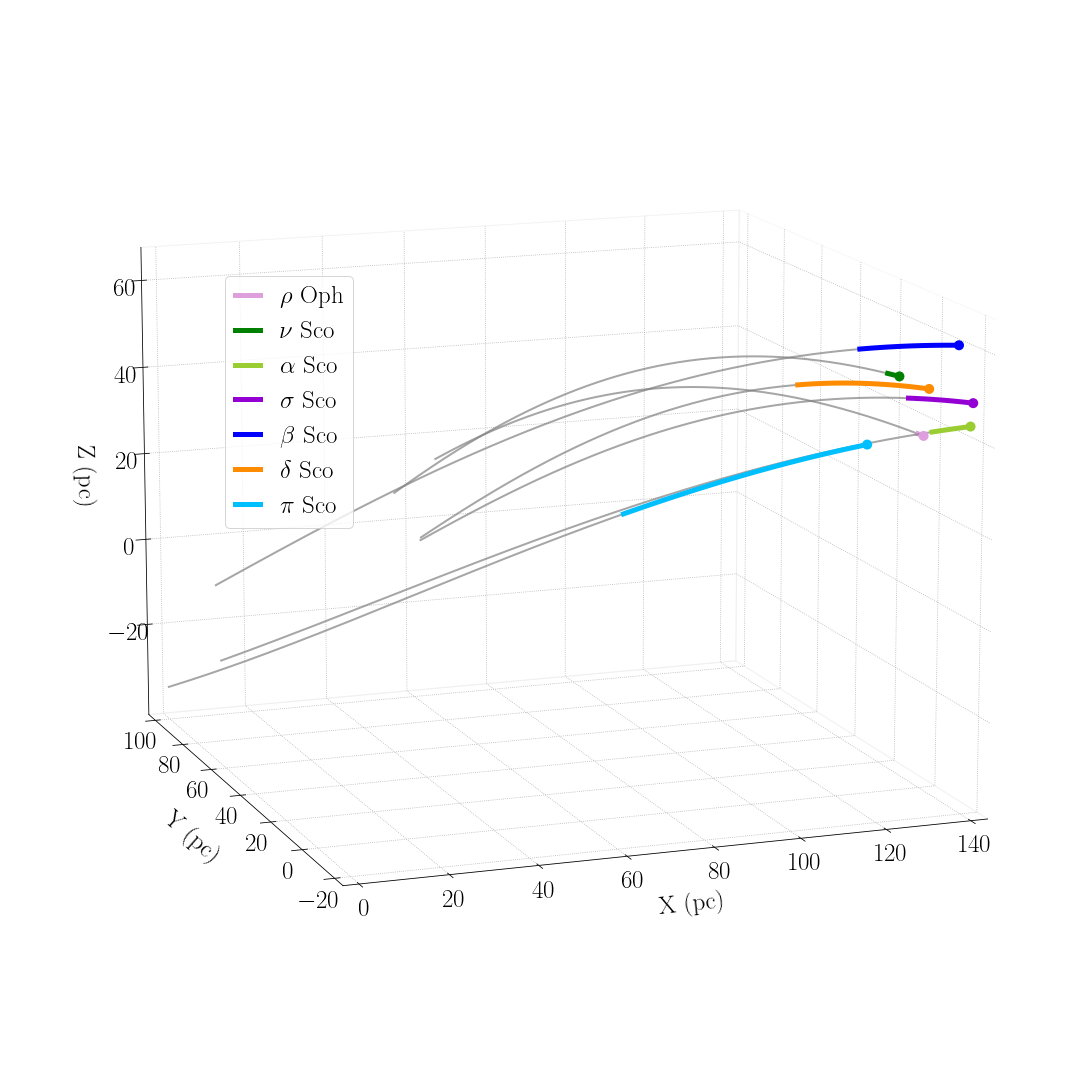}
    \caption{Orbits back in time of the centre of each of the groups identified in this study. The coloured trajectories indicate the orbit of the groups from the present back to the birth time determined in this study within the reported uncertainties in the dynamical ages. The grey trajectories indicate the orbit of the groups from the birth time determined in this study to 20~Myr ago.\\[2.5cm]}
    \label{fig:orbits}
\end{figure*}

We computed the orbit of the centre of each group, 20~Myr back in time using the same strategy described in Sect.~\ref{sec:ages} for individual stars. In Figure~\ref{fig:orbits} (interactive)\footnote{An interactive version of this figure will be public.}, we show the 3D orbits of the centre of each group where the coloured part of the orbit corresponds to the trajectory of the association from the present back to the birth place as determined in this study. 

Upper Scorpius is divided into four different groups according to our analysis namely, $\nu$~Sco, $\sigma$~Sco, $\beta$~Sco and $\delta$~Sco. The groups of $\beta$~Sco and $\delta$~Sco were closest 4~Myr ago, compatible with the dynamical traceback ages of these two groups (2.5$\pm$1.6~Myr and 4.6$\pm$1.1~Myr, respectively), when their centres were at 7~pc. The group of $\nu$~Sco (0.3$\pm$0.5~Myr) was closest to the group of $\beta$~Sco (2.5$\pm$1.6~Myr) 4.5~Myr ago at only 1~pc distance. The orbit of the centre of $\sigma$~Sco (2.1$\pm$0.7~Myr) intersects with $\delta$~Sco (4.6$\pm$1.1~Myr) around 20~Myr ago but these two groups were closer than 10~pc 3~Myr ago. This clearly suggests that these four groups likely have a common origin. 

$\rho$ Oph is the youngest group in our study. We could not estimate a dynamical traceback age since this method requires that the association has undergone a significant expansion and this is not the case for this group. The orbit of the centre of $\rho$~Oph shows that it has a common origin with Upper Scorpius since it was at only 1~pc of $\delta$~Sco 5~Myr ago, similar to the age of this second group. Both the ages obtained in this study, and the fact that $\rho$~Oph is the youngest group in Upper Scorpius, support the formation scenario proposed by \citet{Preibisch+1999}. 

The group $\pi$~Sco is a population located at a distance of 120~pc, in front of Upper Scorpius and Ophiuchus. Our dynamical traceback age (6.3$\pm$1.4~Myr) suggests that it is the oldest group in this study, which is confirmed by the location of the members in this group in a CMD. The group of $\alpha$~Sco has a dynamical traceback age (1.0$\pm$1.2~Myr) similar to the young groups in Upper Scorpius but the CMD indicates that it could be slightly older. These two groups were closest 5~Myr ago when their centres were at 20~pc. Although their orbits seem to cross in the static version of Fig.~\ref{fig:orbits}, they are never at the same place at the same time (see the interactive version). The groups of $\pi$~Sco and $\alpha$~Sco have orbits different from the rest of groups in this study, suggesting that they come from a different origin. We propose that they could have an origin related to other associations in the Sco-Cen complex. We traced back in time the 3D positions and velocities of Upper Centaurus Lupus and Lower Centaurus Crux from \citet{Gagne+2018c} and we found that the groups of $\pi$~Sco and $\alpha$~Sco were closer to Upper Centaurus Lupus 15--20~Myr ago.

\subsection{Stellar feedback}

\begin{table}
\centering
\caption{Strömgren radius. 
\label{tab:stromgren-radius}}
\begin{tabular}{llcccc}
\hline\hline
  \multicolumn{1}{c}{Name} &
  \multicolumn{1}{c}{SpT} & 
  \multicolumn{1}{c}{dist} &
  \multicolumn{1}{c}{log $Q_H$} & 
  \multicolumn{1}{c}{$r_{S, Eq. 3}$} &
  \multicolumn{1}{c}{$r_{S, H\alpha}$} \\
  
  \multicolumn{1}{c}{} &
  \multicolumn{1}{c}{} & 
  \multicolumn{1}{c}{(pc)} & 
  \multicolumn{1}{c}{(s$^{-1}$)} & 
  \multicolumn{1}{c}{(pc)} & 
  \multicolumn{1}{c}{(pc)} \\
\hline\hline
$\delta$ Sco   & B0.3IV (1)     & 150 & 47.99  & 7.9   & 5.5   \\
$\pi$ Sco      & B1V+B2V (2)    & 103 & 47.40  & 5.0   & 1.3   \\
$\sigma$ Sco   & B1III+B1V (3)  & 214 & 47.99  & 7.9   & 2.2   \\
$\tau$ Sco     & B0.2V (4)      & 145 & 47.70  & 6.3   & 3.3   \\
\hline\hline
\end{tabular}
\tablefoot{Columns indicate, (1--3) name of the ionising star, spectral type and distance, (4) hydrogen-ionising photon luminosity (Alves, in prep.), (5) Strömgren radius calculated with Equation~\ref{eq:stromgren_radius}, (6) Strömgren radius measured in H$\alpha$ (see~Fig.~\ref{fig:sky_position} right panel).}
\noindent\tablebib{(1)~\citet{Morgan+1973}; (2)~\citet{Garrison1967}; (3)~\citet{Maiz-Apellaniz+2021}; (4)~\citet{Donati+2006}.}

\end{table}

The orbits discussed in this study are based on the current positions and velocities of stars and a 3D Galactic potential. They do not take into account gravitational interactions between different groups and stellar feedback, which might be important in some cases. In Figure~\ref{fig:sky_position} (right panel), we see the groups identified in this study coinciding with H$\alpha$ emission, tracing the HII regions in this area very likely powered by the four early B-type stars ($\delta$~Sco, $\pi$~Sco, $\sigma$~Sco and $\tau$~Sco). To test whether these HII regions are generated from the photo-ionisation of these massive stars, we compared their size with the radius of Strömgren spheres \citep{Stromgren+1939} defined by the equation: 
\begin{equation}
    r_S = \left(\frac{3~Q_{H}}{4\pi\alpha~n_p^2}\right)^{1/3} 
    \label{eq:stromgren_radius}
\end{equation}
where $Q_H$ is the hydrogen-ionising photon luminosities computed by adopting the calibration in \citet{Smith+2006}, $\alpha = 3.09\cdot10^{-13}$~cm$^3$~s$^{-1}$ is the recombination coefficient \citep[for a kinetic temperature of 8\,000~K, ][]{Spitzer+1978}, and $n_p =$ 10~protons~cm$^{-3}$ is the proton density (Alves, in prep.). The Strömgren radii calculated are listed in Table~\ref{tab:stromgren-radius}.

The sizes of the HII regions measured with Equation~\ref{eq:stromgren_radius} are similar to the measured physical sizes on the H$\alpha$ map, confirming the assumption that photoionization by the current massive stars in the region is responsible for the observed H$\alpha$ emission. Intriguingly, there is no H$\alpha$ emission around the B1V+B2V $\beta$~Sco system. This could indicate that the local density is very low, although more work is needed to unveil this mystery. 
Currently, the groups of $\beta$~Sco and $\nu$~Sco are close but outside the HII region of $\delta$~Sco, suggesting that the present ionisation alone is unlikely to be triggering star formation in these two groups. However, around 4~Myr ago, the centres of the groups $\beta$~Sco and $\nu$~Sco were at the boundaries of the HII region of $\delta$~Sco, suggesting that the photoionization and stellar winds of $\delta$~Sco could have compressed the molecular gas available after the formation of the $\delta$~Sco group, possibly triggering the formation of the groups $\beta$~Sco and $\nu$~Sco.

\begin{figure*}
    \centering
    \includegraphics[width = \textwidth]{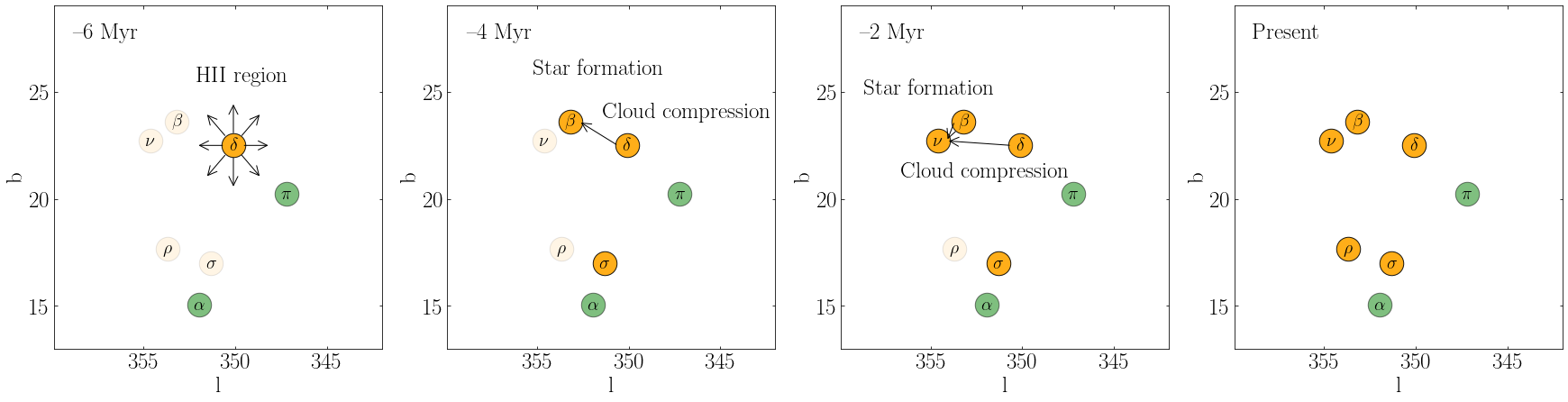}
    \caption{Schema representing the proposed star formation scenario in Upper Scorpius.}
    \label{fig:SF-scenario}
\end{figure*}

In addition to the HII regions surrounding the early B-type stars, we also see other structures in the H$\alpha$ map, such as the large shell at the top image in
Figure~\ref{fig:sky_position} (right panel). This shell could be the remnant of other sources of feedback in the region, although a dedicated analysis goes beyond the goal of this paper. The event that created this shell, not necessarily what currently makes it shine in H$\alpha$, probably had an important impact on the star formation history of the region. About 15 supernovae are known to have exploded in this region \citep{Breitschwerdt+2016, Zucker+2022}, and probably had an important impact on the star formation history of the region. The well-known runaway star $\zeta$~Oph is the smoking gun of a supernovae that exploded about 1--2~Myr ago \citep{Neuhauser+2020}, probably the last supernova explosion in the region. 

In this paper, we analyse the youngest region of Sco-Cen, Upper-Scorpius, which contains about a third of the sources in the entire Sco-Cen OB association. A detailed study of the other neighbouring groups in Sco-Cen will help to shed light on the complete picture, as we have discussed that some groups identified in this study could be older and related to the rest of Sco-Cen. Finally, the large context in which the Sco-Cen association is embedded is relevant to understand the origins of this benchmark association. The Sco-Cen complex is found at the end of the ``Split'', a large-scale gas structure \citep{Lallement+2019}, and next to an even larger gas structure called the ``Radcliffe Wave'' \citep{Alves+2020}, which is likely the gas reservoir of the Orion Arm of the Milky Way \citep{Swiggum+2022}. A full analysis of Sco-Cen using the approach followed in this paper is warranted and is likely to reveal important pieces of the star formation history of the closest OB association, and star formation in the Sun's neighborhood.

\subsection{Proposed star formation scenario}

Taking the orbital traceback analysis as a first-order approximation of the paths of the groups back in time, and complementing this information with the insights on stellar feedback, we propose the following star formation scenario, illustrated in Fig.~\ref{fig:SF-scenario}.

\begin{enumerate}
    \item Around 5--6~Myr ago, the early B-type star $\delta$ Sco ionized the medium around it, creating an HII region.
    \item We propose that in the past, the star-forming cloud extended towards the area currently occupied by the groups $\nu$~Sco and $\beta$~Sco. Then, the combined effect of (possible) supernovae from the $\delta$ Sco group, stellar winds, and the bubble of ionised material around $\delta$~Sco, could have compressed the cloud, triggering star formation mostly towards the area of $\beta$~Sco. 
    \item Around 4~Myr ago, the group of $\beta$~Sco was born. This is the time when it was closest to $\delta$~Sco, which is consistent with the dynamical traceback age determined in this study.
    \item During the last 2~Myr, star formation progressed towards the area of $\nu$~Sco, probably triggered by the feedback of both the $\delta$ Sco and $\beta$ Sco events. Given that there is not much dense gas to the west of $\nu$~Sco, this is perhaps the last significant star formation event in the region. 
\end{enumerate}

This star formation scenario is compatible with a triggered star formation process similar to the one presented by \citet{Preibisch+2008} and with the predictions of the "surround and squash" mechanism proposed by \citet{Krause+2018}. The latter predicts gravitationally unbound groups with substructures that have coherent kinematics, as we observed in our sample.

\section{Conclusions}\label{sec:conclusions}

We present the first detailed study of the star formation history of Upper Scorpius and Ophiuchus in 7D (3D positions, 3D velocities and age). Combining \textit{Gaia} DR3 astrometry and radial velocities with precise ground-based radial velocities from APOGEE DR17 and our own observations, we identified seven groups using 3D kinematics and 3D spatial information. We integrated the orbits of individual stars back in time with a 3D Galactic potential and measured dynamical traceback ages (i.e., the time when the stars in a group were most concentrated in the past, \citealt{Miret-Roig+18, Miret-Roig+2020b}). We found an age gradient between different groups that is statistically significant and that also correlates with an age gradient in colour-magnitude diagrams. We found a dynamical traceback age of Upper Scorpius younger than 5~Myr.

We studied the star formation history of this complex by computing the orbit of the centre of each group back in time. We found that the Upper Scorpius association splits into four groups namely, $\nu$~Sco, $\beta$~Sco, $\sigma$~Sco and $\delta$~Sco, which share a common origin. We speculate that feedback from  massive stars in the $\delta$~Sco group could have triggered star formation first in $\beta$~Sco and then in $\nu$~Sco. The origin of $\rho$~Oph can also be related to Upper Scorpius according to its orbit back in time, in agreement with what previous studies proposed \citep{Preibisch+1999, Preibisch+2002}. The two groups of $\alpha$~Sco and $\pi$~Sco seem to have a different origin, which we relate to other regions of the Sco-Cen complex (Upper Centaurus-Lupus). 

The star formation history presented in this study constitutes an advance from the early work of \cite{Preibisch+1999}. We confirm their global picture but we show that the history of the region is much more complex than in their scenario. Still, the final picture is far from complete since a detailed analysis of stellar feedback and gas dynamics is missing, which will have an important impact on the large-scale gas motion of the region. 

\begin{acknowledgements}
The authors thank the anonymous referee for very useful comments that helped to improve the presentation of these results. 
The authors thank Cameren Swiggum for producing the interactive version of Figure~\ref{fig:orbits}. 
This research has received funding from the European Research Council (ERC) under the European Union’s Horizon 2020 research and innovation programme (grant agreement No 682903, P.I. H. Bouy), and from the French State in the framework of the ”Investments for the future” Program, IdEx Bordeaux, reference ANR-10-IDEX-03-02. We thank NVIDIA Corporation for their support, in particular, the donation of one of the Titan Xp GPUs used for this research. 
P.A.B. Galli acknowledges financial support from São Paulo Research Foundation (FAPESP) under grants 2020/12518-8 and 2021/11778-9.
JO acknowledges financial support from “Ayudas para contratos postdoctorales de investigación UNED 2021”.
DB has been partially funded by MCIN/AEI/10.13039/501100011033 grants PID2019-107061GB-C61 and MDM-2017-0737.
This work has made use of data from the European Space Agency (ESA) mission {\it Gaia} (\url{https://www.cosmos.esa.int/gaia}), processed by the {\it Gaia} Data Processing and Analysis Consortium (DPAC, \url{https://www.cosmos.esa.int/web/gaia/dpac/consortium}). Funding for the DPAC has been provided by national institutions, in particular the institutions participating in the {\it Gaia} Multilateral Agreement.
Based on observations collected at the European Organisation for Astronomical Research in the Southern Hemisphere under ESO programmes 
099.A-9029, 083.D-0034, 083.A-9013, 083.A-9017, 093.A-9029, 077.C-0258, 076.A-9018, 079.A-9002, 072.A-9012, 087.C-0315, 084.A-9016, 081.C-2003, 073.C-0355, 094.A-9012, 093.A-9008, 073.C-0337, 086.D-0449, 082.D-0061, 087.D-0099, 089.D-0153, 093.A-9006, 083.A-9003, 084.D-0067, 179.C-0197, 077.D-0477, 091.C-0713, 076.D-0172, 075.C-0399, 072.A-9006, 078.C-0011, 077.C-0138, 092.A-9029, 091.C-0216, 178.D-0361, 0101.A-9003,
098.C-0739, 082.C-0390, 081.C-0779, 077.D-0085, 099.D-0380, 0104.C-0418,0100.D-0273,1101.C-0557,
099.D-0628, 188.B-3002, 081.C-0708, 075.C-0256, 073.C-0179, 075.B-0863, 075.C-0551,
65.I-0404,  69.C-0481,  65.L-0199,  67.C-0160,  69.B-0108,  71.C-0068,  097.C-0409, 084.C-1002, 079.C-0556, 083.D-0689, 266.D-5655, 073.C-0138, 081.C-0475, 077.C-0323, 093.C-0476, 085.C-0524, 093.C-0658, 075.C-0272, 097.C-0979, 082.C-0005, 081.D-0904, 079.C-0375, 093.D-0279, 081.C-0222, 075.C-0292, 089.C-0299, 0102.C-0040.
Based on spectral data retrieved from the ELODIE archive at Observatoire de Haute-Provence (OHP, http://atlas.obs-hp.fr/elodie/). Based on observations obtained withCHIRON at the 1.5m Telescope in Cerro Tololo operated by the SMARTS Consortium (P.I. Bouy, NOIRLab Programs 2020A-0094 and 2021A-0011) with Hermes at the Mercatortelescope (P.I. Barrado, Program 5-Mercator1/21A).
\end{acknowledgements}

\begin{appendix}

\section{Additional Tables and Figures}

\begin{figure*}
    \centering
    \includegraphics[width = 0.7\textwidth]{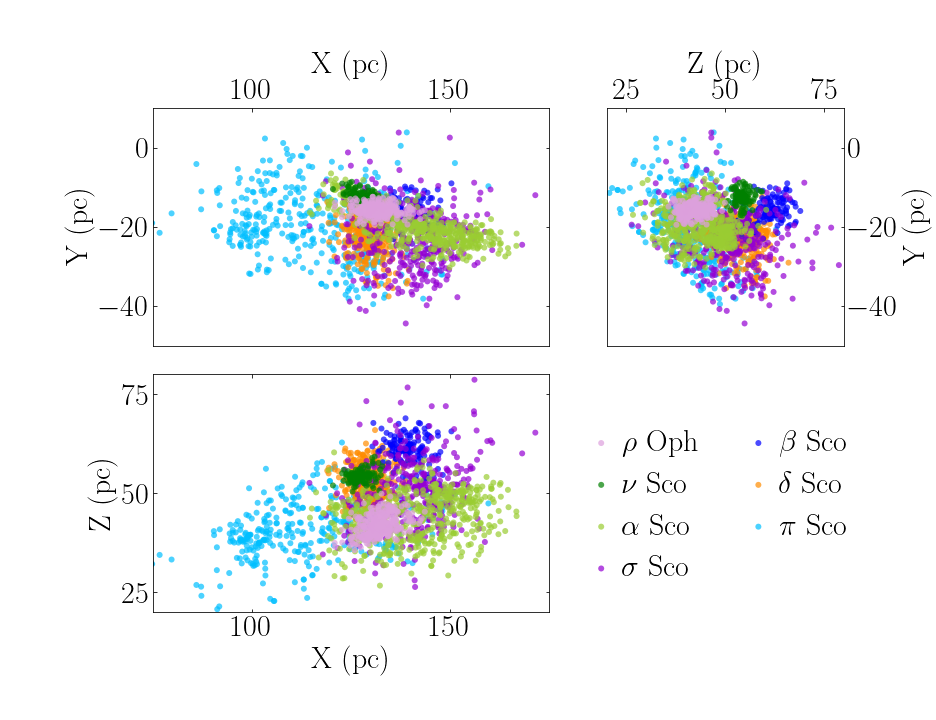}
    \includegraphics[width = 0.6\textwidth]{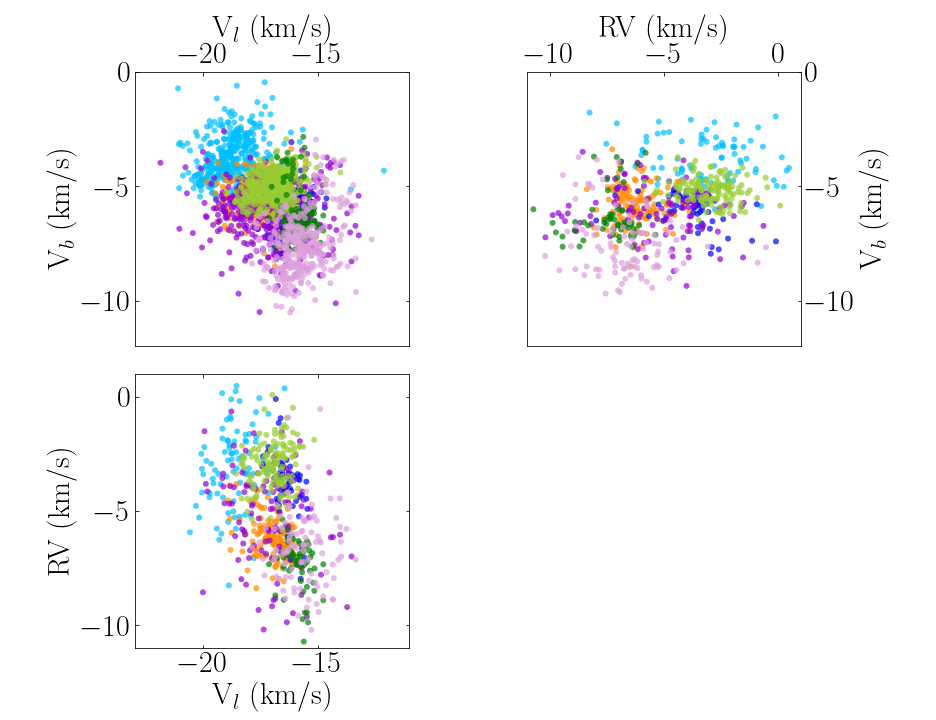}
    \caption{Distribution of the groups in the representation space. }
    \label{fig:rs}
\end{figure*}

\begin{table*}
\setlength{\tabcolsep}{4pt}
\centering
\caption{Comparison with previous studies. 
\label{tab:literature}}
\begin{tabular}{l|ccc} 
\hline\hline

  \multicolumn{1}{c|}{Group name}  & 
  \multicolumn{1}{c}{\citet{Kerr+2021}} & 
  \multicolumn{1}{c}{\citet{Squicciarini+2021}} & 
  \multicolumn{1}{c}{\citet{Ratzenbock+2022}} \\
 
\hline\hline
$\rho$ Oph   & LEAF I       &   i Sco        &  $\rho$ Oph   \\ 
$\nu$ Sco    & LEAF E       &   $\nu$ Sco B  &  $\nu$ Sco    \\ 
$\alpha$ Sco & LEAF N $^+$  &   diffuse      &  Antares, $\sigma$ Sco   \\ 
$\sigma$ Sco & LEAF N $^+$  &   diffuse      &  $\sigma$ Sco $^+$    \\ 
$\beta$ Sco  & LEAF G       &   HD 144273    &  HD 146367    \\ 
$\delta$ Sco & LEAF H       &   b Sco        &  $\delta$ Sco   \\ 
$\pi$ Sco    & LEAF N $^+$  &   diffuse      &  HD 145964 $^+$   \\ 

\hline\hline
\end{tabular}
\tablefoot{Our group $\alpha$ Sco also coincides with Pop 2 in \citet{Grasser+2021}.\\ $^+$ Include a significant fraction of other groups.}

\end{table*}

\begin{figure*}
    \centering
    \includegraphics[width = 0.9\textwidth]{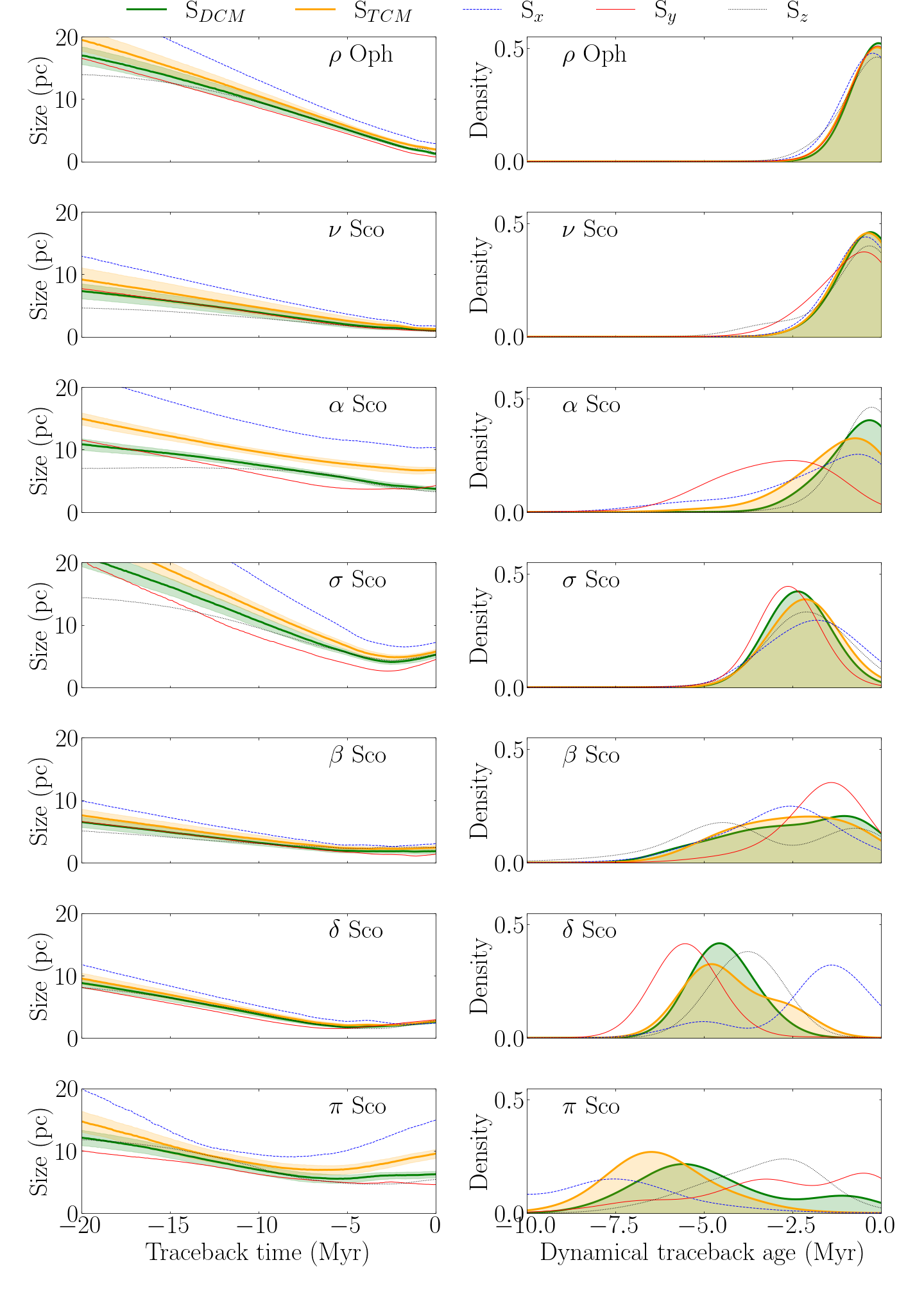}
    \caption{Dynamical traceback ages. Left panels show the size of each group as a function of traceback time for the different size functions defined in Sect.~\ref{sec:ages}. Right panels show the dynamical traceback ages for the different size functions.}
    \label{fig:dyn_age_size_time}
\end{figure*}

\end{appendix}

\bibliographystyle{aa} 
\bibliography{mybiblio} 

\end{document}